\input harvmac

\Title{\vbox{\rightline{EFI-98-40}\rightline{hep-th/9809021}}}
{\vbox{\centerline{Conformal Field Theory, Geometry, and Entropy}}}
\bigskip

\baselineskip=12pt
\centerline{Emil J. Martinec%
\footnote{$^\dagger$}{\tt e-martinec@uchicago.edu}} 
\bigskip
\centerline{\sl Enrico Fermi Inst. and Dept. of Physics}
\centerline{\sl University of Chicago}
\centerline{\sl 5640 S. Ellis Ave., Chicago, IL 60637, USA}

\baselineskip=16pt
 
\vskip 2cm
\noindent

In the context of the AdS/CFT correspondence,
an explicit relation between the physical degrees of freedom 
of 2+1d gravity and the stress tensor
of 1+1d conformal field theory is exhibited.
Gravity encodes thermodynamic state variables of
conformal field theory, but does not distinguish among different
CFT states with the same expectation value for the stress tensor.
Simply put, gravity is thermodynamics; gauge theory is
statistical mechanics.

\Date{9/98}

%
%
\def\journal#1&#2(#3){\unskip, \sl #1\ \bf #2 \rm(19#3) }
\def\andjournal#1&#2(#3){\sl #1~\bf #2 \rm (19#3) }

\def\ie{{\it i.e.}}
\def\eg{{\it e.g.}}
\def\cf{{\it c.f.}}

\def\etc{{\it etc.}}

\def\sst{\scriptscriptstyle}

\def\frac#1#2{{#1\over#2}}
\def\coeff#1#2{{\textstyle{#1\over #2}}}
\def\half{\frac12}
\def\hf{{\textstyle\half}}
\def\ket#1{|#1\rangle}

\def\vev#1{\langle#1\rangle}
\def\d{\partial}

\def\inbar{\,\vrule height1.5ex width.4pt depth0pt}
\def\IC{\relax\hbox{$\inbar\kern-.3em{\rm C}$}}
\def\IR{\relax{\rm I\kern-.18em R}}
\def\IP{\relax{\rm I\kern-.18em P}}
\def\IH{\relax{\rm I\kern-.18em H}}

\def\nth{$n^{\rm th}$}
%
%

\def\npb#1#2#3{Nucl. Phys. {\bf B#1} (#2) #3}

\def\plb#1#2#3{Phys. Lett. {\bf #1B} (#2) #3}
\def\prl#1#2#3{Phys. Rev. Lett. {\bf #1} (#2) #3}

\def\prd#1#2#3{Phys. Rev. {\bf D#1} (#2) #3}

\def\cmp#1#2#3{Comm. Math. Phys. {\bf #1} (#2) #3}
\def\cqg#1#2#3{Class. Quant. Grav. {\bf #1} (#2) #3}

\def\jhep#1#2#3{J. High Energy Phys. {\bf #1} (#2) #3}
\catcode`\@=11
\def\slash#1{\mathord{\mathpalette\c@ncel{#1}}}
\overfullrule=0pt
\def\AA{{\cal A}}

\def\FF{{\cal F}}
\def\GG{{\cal G}}

\def\II{{\cal I}}

\def\MM{{\cal M}}

\def\OO{{\cal O}}

\def\RR{{\cal R}}

\def\ZZ{{\cal Z}}
\def\lam{\lambda}

\def\underrel#1\over#2{\mathrel{\mathop{\kern\z@#1}\limits_{#2}}}

\catcode`\@=12


%

\def\ket#1{\left| #1\right\rangle}
\def\vev#1{\left\langle #1 \right\rangle}

\def\Tr{{\rm Tr}}

\def\cosh{{\rm cosh}}

\def\exp{{\rm exp}}


\def\cft{{\sst\rm CFT}}
\def\st{\scriptstyle}
\def\lpl{\ell_{{\rm pl}}}

\def\sltwo{{$SL(2,R)$}}
\def\lz{{L_0}}
\def\lzb{{{\tilde L}_0}}
\def\atil{{\tilde A}}

\def\zbar{{\bar z}}
\def\ee{{e}}
\def\oo{{\omega}}
\def\liou{{\sst\rm liou}}
%
\nref\maldaconj{J. Maldacena, hep-th/9711200.}%
\nref\gkp{S.S. Gubser, I.R. Klebanov, and A.M. Polyakov, hep-th/9802109.}%
\nref\wittenads{E. Witten, hep-th/9802150; hep-th/9803131.}%
\nref\bdhm{T. Banks, M. Douglas, G. Horowitz, and E. Martinec,
hep-th/9808016.}%
\nref\bklt{V. Balasubramanian, P. Kraus, A. Lawrence, and S. Trivedi, 
hep-th/9808017.}%
\nref\susswit{L. Susskind and E. Witten, hep-th/9805114.}%
\nref\btz{M. Ba\~nados, C. Teitelboim, and J. Zanelli,
\prl{69}{1992}{1849}.  For an extensive review and further
references, see S. Carlip, gr-qc/9506079;
\cqg{12}{1995}{2853}.}%
\nref\brownhenn{J. Brown and M. Henneaux, \cmp {104}{1986}{207}.}%
\nref\strom{A. Strominger, hep-th/9712251.}%
\nref\maldastrom{J. Maldacena and A. Strominger, hep-th/9804085.}%
\nref\adsmm{E. Martinec,  hep-th/9804111; see also
http://www.itp.ucsb.edu/online/dual/martinec1.}%
\nref\cardy{J. Cardy, \npb{270}{1986}{186}.}%
\nref\jacobsen{T. Jacobsen, gr-qc/9504004; \prl{75}{1995}{1260}.}%
\nref\carlipent{S. Carlip, gr-qc/9409052; \prd{51}{1995}{632}.}
\nref\dunno{S. Carlip, hep-th/9806026.}%
\nref\hipt{A. Achucarro and P.K. Townsend, \plb{180}{1986}{89};
P.S. Howe, J.M. Izquierdo, G. Papadopoulos, and
P.K. Townsend, hep-th/9505032.}%
\nref\witcsgrav{E. Witten, \npb{311}{1988}{46}; \npb{323}{1989}{113}.}%
\nref\dodf{A. Strominger and C. Vafa, hep-th/9601029;
\plb{379}{1996}{99};
J. Maldacena, Princeton Ph.D. thesis, hep-th/9607235,
see also hep-th/9705078, Nucl. Phys. Proc. Suppl. {\bf 61A} (1998) 111-123;
R. Dijkgraaf, E. Verlinde, and H. Verlinde,
hep-th/9704018, \npb{506}{1997}{121};
J. Maldacena, A. Strominger, and E. Witten, hep-th/9711053.}%
\nref\chvd{O. Coussaert, M. Henneaux, and P. van Driel,
gr-qc/9506019; \cqg {12}{1995}{2961}.}%
\nref\aleks{A. Alekseev and S. Shatashvili, \npb{323}{1989}{719};
M. Bershadsky and H. Ooguri, \cmp{126}{1989}{49}.}%
\nref\banados{M. Ba\~nados, hep-th/9405171; \prd {52}{1996}{5816}.}%
\nref\bbo{M. Ba\~nados, T. Brotz, and M. Ortiz, hep-th/9802076.}%
\nref\bkl{V. Balasubramanian, P. Kraus, and A. Lawrence, hep-th/9805171.}%
\nref\nav{J. Navarro-Salas and P. Navarro, hep-th/9807019.}%
\nref\cz{N. Cruz and J. Zanelli, gr-qc/9411032; \cqg{12}{1995}{975}.}%
\nref\deboer{J. deBoer, hep-th/9806104.}%
\nref\gibhawk{G. Gibbons and S. Hawking, \cmp{66}{1979}{291}.}%
\nref\carteit{S. Carlip and C. Teitelboim, gr-qc/9405070,
\prd{51}{1995}{622}; S. Carlip, gr-qc/9606043, \prd{55}{1997}{878}.}%
\nref\banmen{M. Banados and F. Mendez, hep-th/9806065.}%
\nref\dansteve{D. Friedan and S. Shenker, \npb{281}{1987}{509}.}%
\nref\kutseib{D. Kutasov and N. Seiberg, \npb{358}{1991}{600}.}%
\nref\thorn{C. Thorn, \npb{248}{1984}{551}.}%
\nref\tangle{L. Bombelli, R.K. Koul, J. Lee, and R.D. Sorkin,
\prd{34}{1986}{373}.; V. Frolov and I. Novikov, \prd{48}{1993}{4545};
M. Srednicki, hep-th/9303048, \prl{71}{1993}{666}.}%
\nref\fpst{T. Fiola, J. Preskill, A. Strominger, and S. Trivedi,
hep-th/9403137; \prd{50}{1994}{3987}.}%
\nref\matbh{T. Banks, W. Fischler, I. Klebanov, and L. Susskind,
hep-th/9709091, \prl{80}{1998}{226}; hep-th/9711005,
\jhep{1}{1998}{8}; G. Horowitz and E. Martinec, hep-th/9710217,
\prd{57}{1998}{4935}.}%
\nref\stromtalk{A. Strominger, talk at Strings'98,
http://www.itp.ucsb.edu/online/strings98/strominger.}
%

\newsec{Introduction}

The thermodynamic character of gravity in the presence of black holes
has led to a longstanding search for an underlying statistical mechanics.  
This quest has been plagued by a number of conceptual issues. 
In a local field theoretic approach to gravity,
what is the meaning of equilibrium thermodynamics of black holes
when macroscopic regions of the spacetime are out of causal contact --
especially since the origin of the thermodynamics
is the presence of the horizon itself?
If one can localize field theoretic excitations in a finite region,
where and of what nature are the set of field configurations
that characterize the entropy?

While a definitive answer to these questions has not yet been found,
the duality between anti-de Sitter (super)gravity and 
conformal field theory (CFT) conjectured \maldaconj\
on the basis of recent advances in string theory 
may contain the key physical insights.
The maximal scope of the conjecture posits
that the full M/string theory in asymptotically anti-de Sitter
spacetimes $AdS_p\times K$
is equivalent to a particular conformal field theory in $p-1$ 
spacetime dimensions.
In this construction, a conventional quantum field theory
(with positive norm Hilbert space and unitary evolution) --
which is the infrared limit of some generalized gauge theory
of brane dynamics --
provides the underlying degrees of freedom.  
``Local'' quantum fields coupled to gravity 
are highly composite operators 
\refs{\gkp-\bdhm}
built from these gauge theory degrees of freedom;
however, the extent to which excitations may be
actually localized in $AdS_p\times K$ is not clear.
The fact that the density of states of the dual description
grows asymptotically like that of a lower
dimensional field theory (rather than a ten- or eleven-dimensional
M/string theory) militates against locality \susswit.
On the other hand, this density of states 
is compatible with the entropy of AdS-Schwarzschild
black holes \refs{\wittenads,\susswit}.

Gravity in 2+1 dimensions is a useful arena for the exploration
of the relation between quantum black holes and thermodynamics.
In the presence of a negative cosmological constant 
$\Lambda=-\frac 1{\ell^2}$, 
2+1 gravity admits the analogue of anti-de Sitter Schwarzschild black hole
solutions \btz, known as BTZ black holes.  
These solutions exhibit all the usual thermodynamic
properties of black holes: their entropy is the horizon area in Planck units
\eqn\btzent{
  S=\frac{2\pi r_+}{4G}\ ,
}
and they obey the first law 
\eqn\firstlaw{
  dE=TdS+\Omega dJ\ .
}
Here $E$ is the ADM energy,
$T$ is the Hawking temperature (defined as $1/2\pi$ times the
surface gravity $\kappa$), $J$ is the angular momentum, and $\Omega$
is the angular potential.

A useful property of pure 2+1 gravity is the absence of dynamical
bulk degrees of freedom; the gauge freedom is sufficient to 
push all the dynamics onto the boundaries of the space --
horizons, naked singularities (such as are produced by heavy point
particle sources), and the timelike boundary at spatial infinity
when $\Lambda<0$.  Thus, there is a clean separation between
the gravitational sector, containing only global degrees of freedom;
and any given matter sector, whose local bulk dynamics one
wishes to couple to gravity.

For these reasons, we wish to concentrate on
the particular example of $AdS_3$ gravity, because all the ingredients
of the black hole puzzle are present in a very controlled
setting.  Both $AdS_3$ gravity and 
its proposed dual 1+1 dimensional conformal field theory 
are representations of the infinite-dimensional
Virasoro algebra with central extension 
\refs{\brownhenn-\adsmm}
\eqn\centext{
  c=\frac{3\ell}{2G}\ .
}
The unitary CFT has an asymptotic density of states \cardy\
\eqn\levdens{
  S=2\pi[(cL_0/6)^{\half}+(c\lzb/6)^\half]\ ;
}
this level density matches that of BTZ black holes \btz\ with
mass and spin determined by equating the 
Casimirs of the Virasoro representation
\eqn\masspin{
  \ell M=\lz+\lzb\qquad,\quad\qquad J=\lz-\lzb\ .
}
Thus one expects that the states of the CFT are
indeed the microstates responsible for the BTZ black hole entropy.

The point of view we will try to justify here is that 2+1 gravity
is a collective field excitation of the underlying 
dual conformal field theory description \adsmm, constructed from
the CFT stress tensor.   
Because gravity itself carries no local excitations,
only the global geometric data of the spacetime 
appears in the construction.
It is this global data (which can be thought of as a 
set of Noether charges \brownhenn)
that couples to thermodynamics.  On the other hand,
since it is constructed solely from the CFT stress tensor,
gravity cannot distinguish among CFT states of the same
energy and other charges.  
Indeed, as we shall see, the density of states of 
gravity is \levdens\ with $c=c_{\rm eff}=1$ rather than \centext.
To summarize the situation in brief:

\bigskip
\item{ }
{\it Gravity is thermodynamics; gauge theory (of branes)
is statistical mechanics}.

\bigskip
\noindent
Since the classical gravity solution represents the typical 
microstate of the underlying CFT with the same
global properties (or equivalently, an average over
these microstates), one should not be asking gravity to 
provide an explanation of the entropy.
It is quite possible that gravity itself will ultimately
be understood as a thermodynamic phenomenon; for thoughts
along these lines, see \jacobsen.

\newsec{Gravity on $AdS_3\times K$ and its CFT dual}

We will focus for the moment on the pure gravity sector.%
\foot{There are claims in the literature 
\refs{\carlipent} that this is already sufficient to explain 
2+1d black hole entropy (although see \refs{\adsmm,\dunno}).}
In general, one is interested in gravity coupled to matter
(a particular case of interest is the matter arising from
string theory compactification).
Later we will return to the 
the inclusion of matter into our considerations.
The low-energy regime of gravity in any of these theories
is described by $AdS_3$ Chern-Simons (super)gravity
\refs{\hipt,\witcsgrav}
with gauge group $SL(2,R)_L\times  SL(2,R)_R$,%
\foot{The supersymmetric completion is irrelevant for the 
issues arising in the present investigation.}
which is the global conformal symmetry of the dual CFT.%
\foot{In the particular case of bound states of D1- and D5-branes,
$K=S^3\times M$, and a candidate for the dual CFT is
a resolution of $S^k(M)$ -- 
the symmetric orbifold of $k=Q_1Q_5$ copies of $M$ \dodf.}
The convenient variables are the `gauge'
fields $A=\oo+\ee/\ell$ and $\tilde A=\oo-\ee/\ell$,
in terms of which the action is
\eqn\threedgrav{
  \frac1{16\pi G}\int \; \ee(\RR-2\Lambda)=\frac{k}{4\pi}\Bigl(
        \int(AdA+\coeff23 A^3)-\int(\atil d\atil+\coeff23\atil^3)\Bigr)\ .
}
Here $\Lambda=-\ell^{-2}$, and $k=\ell/(4G)$; $\ell$ is the
$AdS_3$ radius, and $G\sim \lpl$ is the 2+1d Planck scale.

It was shown by Brown and Henneaux \brownhenn\ that the global
conformal algebra extends to the full Virasoro algebra
of diffeomorphisms which preserve the asymptotically anti-de Sitter
form of the metric; the resulting
algebra of gravitational Noether charges $L_n$, ${\bar L}_n$
has central charge $c=6k=\frac{3\ell}{2G}$.  
In global coordinates where the metric takes the asymptotic form
\eqn\asympmet{
  ds^2\sim 
	\ell^2\Bigl(\frac{1}{r^2}dr^2
	-r^2 du\,dv 
	+\gamma_{uu}(du)^2 +\gamma_{vv}(dv)^2
	+\OO(1/r)\Bigr)\ ,
}
these diffeomorphisms are analytic reparametrizations
of $u$ and $v$ ($\theta=\half(u-v)$ is taken periodic:
$\theta\sim\theta+ 2\pi$).  
Here and below, all coordinates will be made dimensionless
by referring them to the scale $\ell$.
Declaring that these diffeomorphisms
are not allowed gauge symmetries imposes the boundary
conditions that the connections $A$, $\atil$
are asympototically pure gauge $A_\mu\sim \GG^{-1}\d_\mu \GG$, 
$\atil_\mu\sim \d_\mu {\tilde \GG}\cdot{\tilde \GG}^{-1}$, with
\eqn\pure{\eqalign{
  \GG=&~\pmatrix{\sqrt{r}&0\cr 0&\frac{1}{\sqrt{r}}} g(u)\cr
  {\tilde \GG}=&~{\tilde g}(v)\pmatrix{\frac1{\sqrt{r}}&0\cr 0&\sqrt{r}}\ ;
}}
furthermore, $g(u)$ and ${\tilde g}(v)$ undergo
a Borel-type (\ie, upper- or lower-triangular)
restriction \chvd\ that reduces the corresponding affine
$SL(2,R)$ currents $g^{-1}\d_u g$, $\d_v g\cdot g^{-1}$
to left and right Virasoro algebras
with $c=6k$ \aleks.  
Canonical expressions for the generating
functions of these algebras are ($\alpha=\hf\sigma_3$)
\refs{\banados,\bbo}
\eqn\cangen{\eqalign{
  T_{uu}=\sum_n L_n e^{-inu}=&~ \frac{k}{2}{\rm Tr}\bigl(
        2\alpha\d_u A_u + A_uA_u\bigr)\cr
  T_{vv}=\sum_n L_n e^{-inv}=&~ \frac{k}{2}{\rm Tr}\bigl(
        2\alpha\d_v \atil_v + \atil_v\atil_v\bigr)\ .
}}

The trivial bulk dynamics of pure 2+1 gravity allows one
to collapse its dynamics onto the boundary; one finds \chvd\
a Liouville action, and the generators \cangen\
are just the components of the Liouville stress tensor
\eqn\listress{\eqalign{
  T_{uu}^\liou=&~ k(\d_u\varphi\d_u\varphi - \d_u^2\varphi)
	=k\gamma_{uu}\cr
  T_{vv}^\liou=&~ k(\d_v\varphi\d_v\varphi - \d_v^2\varphi)
	=k\gamma_{vv}\ .
}}
In principle, there may be several boundaries of the space:
Timelike singularities corresponding to point particle sources;
the timelike boundary at spatial infinity of anti-de Sitter space;
and one may also wish to consider the horizon 
of a black hole as a boundary.
The first two of these have sensible interpretations
in the AdS/CFT correspondence, whereas the last does not; 
below, we will try to argue 
that the black hole horizon should not be taken as
a boundary of spacetime.

In the AdS/CFT correspondence, 
the CFT dual to 2+1 AdS gravity is a representation
of the same algebra of global symmetries, generated
by the CFT stress tensor.
At the semiclassical level, 
the symmetry generators of the two theories must match.
Thus one is led to propose the identification of the Liouville
field $\varphi$ as a kind of collective coordinate
of the dual CFT at large $k$ via
\eqn\collect{
  \vev{T_{uu}}_{\cft}=T_{uu}^{\liou}
}
This identification is an example of the general correspondence
\refs{\bkl,\nav,\bklt}
between the subleading asymptotic behavior of a bulk field
and the expectation value of an operator in the CFT.
The relation \collect\
is not meant to imply that the underlying large $k$ CFT
is entirely equivalent to Liouville theory; rather, we
are simply making use of the fact that the current
sector of any conformal field theory is universal,
and that the Liouville action is the universal effective action
that encodes the Virasoro Ward identities at the semiclassical level.
One expects that the quantum fluctuations of
the two theories are rather different.

Classical solutions to 2+1 gravity can be characterized by
the holonomies of the two Chern-Simons connections $A$, $\atil$.
Recall that holonomies in \sltwo\ fall into three conjugacy classes:
\item{1.} Hyperbolic elements, conjugate to a dilation
$g\sim \pmatrix{e^{2\pi\lam}&0\cr0&e^{-2\pi\lam}}$.
\item{2.} Parabolic elements, conjugate to a translation
$g\sim \pmatrix{1&2\pi a\cr0&1}$.
\item{3.} Elliptic elements, conjugate to a rotation
$g\sim \pmatrix{\ \cos\,2\pi\alpha&\sin\,2\pi\alpha\cr 
-\sin\,2\pi\alpha&\cos\,2\pi\alpha}$.

\medskip
\noindent 
Let us first consider several examples; then we will re-examine
these solutions in the framework of Liouville/Virasoro theory.

\bigskip

\noindent{{\it `Particle' states:}}
Classical particle sources in 2+1 gravity introduce
conical timelike singularities in the geometry, with the 
deficit angle $\pi(\alpha+\tilde\alpha)$
proportional to the mass of the particle.  
Around such sources, the connections $A$ and $\atil$ 
have elliptic holonomy $cos[2\pi\alpha]$
and $cos[2\pi\tilde\alpha]$, respectively.
The exact classical solution for a single source is
\eqn\ptcleconn{\eqalign{
  A=&~\hf\pmatrix{\frac{dr}{(r^2+\gamma^2)^{1/2}} & 
		{\st (r+(r^2+\gamma^2)^{1/2}) du} \cr
	{\st (r-(r^2+\gamma^2)^{1/2}) du} &
		-\frac{dr}{(r^2+\gamma^2)^{1/2}} \cr}   \cr
  \atil=&~\hf\pmatrix{-\frac{dr}{(r^2+{\tilde\gamma}^2)^{1/2}} & 
		{\st (r-(r^2+{\tilde\gamma}^2)^{1/2}) dv} \cr
	{\st (r+(r^2+{\tilde\gamma}^2)^{1/2}) dv} &
		\frac{dr}{(r^2+{\tilde\gamma}^2)^{1/2}} \cr} \ ,  \cr
}}
where $\gamma=1-\alpha$, and $\tilde\gamma=1-\tilde\alpha$.
Such sources can be thought of as Wilson lines of the Chern-Simons 
gravity theory.  Particles with spin have $\alpha\ne\tilde\alpha$.
The connection \ptcleconn\ leads to a stress tensor \cangen\
with $\lz=-\frac k2 \gamma^2$; the vacuum $AdS_3$
corresponds to $\gamma=1$.  A geometrical particle
source adds energy $\lz=-\frac k2 \alpha(\alpha-1)$ 
above the $AdS_3$ vacuum; its ADM mass and spin are
\eqn\adsenspin{\eqalign{
	\ell M=&~ L_0+\lzb -M_0 \cr
	J=&~ L_0-\lzb 
}}
relative to the $AdS_3$ vacuum of `mass' $\ell M_0=-k/2$ and spin $J=0$.

One can compare this geometrical source with the quanta
of a massive field in $AdS_3$.
The field solves the wave equation
\eqn\waveq{
  \Delta=-(L_1L_{-1}+L_{-1}L_1)+2L_0^2=\frac{m^2\ell^2}2\ .
}
The eigenfunctions of the wave operator with a given mass
form a discrete series representation of $SL(2,R)_L\times SL(2,R)_R$,
with highest weight
\eqn\scalarmass{
  L_0=\lzb=\hf[1+(m^2\ell^2+1)^{\half}]\ .
}
For sufficiently large mass $m\ell\gg 1$ -- so that the 
particle can be treated as a semiclassical source for its gravitational
field -- the particle created by this field 
adds an ADM energy $\delta M=1+(m^2\ell^2+1)^{\half}\sim m\ell$
to the $AdS$ vacuum.  Thus we see that, to leading order,
the geometrical notion of mass in 2+1 gravity agrees with 
the kinematic mass in the wave equation: $m\ell\sim k\alpha$.

\bigskip
\noindent{{\it The 2+1 black hole:}}
The connections are \btz,\bbo\ ($\rho$ is a radial coordinate,
asympotic to $log(r)$)
\eqn\bhconn{
  A=\hf\pmatrix{d\rho & z_+e^\rho du\cr z_+e^{-\rho}du & d\rho}
	\qquad\qquad
  \atil=\hf\pmatrix{d\rho & z_-e^{-\rho} dv\cr z_-e^{\rho}dv & d\rho}\ ,
}
in terms of which the mass and spin are
\eqn\bhenspin{\eqalign{
        \ell M=&~ L_0+\lzb = k(z_+^2+z_-^2)\cr
        J=&~ L_0-\lzb = k(z_+^2-z_-^2)\ ;
}}
Geometrically, $r_\pm=\half(z_+\pm z_-)$ are the radii of 
the inner and outer horizons.
The holonomies at constant $\rho$, $t$ are
\eqn\bhholon{
  \Tr(\exp\oint A)=2\,\cosh(\pi z_+)
	\quad,\qquad
  \Tr(\exp\oint\atil)=2\,\cosh(\pi z_-)\ .
}
These holonomies are in the hyperbolic conjugacy class of
\sltwo\ (similarly for $exp[\oint\atil]$).
The extremal limit arises by taking, say, $z_-$ to zero;
then $exp[\oint A]$ is hyperbolic, and $exp[\oint\atil]$
is parabolic.  The `double extreme' black hole,
with $M=J=0$, has $z_+=z_-=0$, and both holonomies
are parabolic; this solution is also reached as the limit of
the particle solution \ptcleconn\ where
$\gamma,\tilde\gamma\rightarrow 0$.
A solution with nontrivial expectation value for the $L_n$'s
is the exact `gravitational wave' solution built by a
simple modification of the double extreme black hole
\eqn\adswave{
  A\sim\hf\pmatrix{dr/r&rdu\cr (1/kr)T_{uu}du&-dr/r}
	\qquad\qquad
  \atil\sim\hf\pmatrix{-dr/r&0\cr rdv&dr/r}\ .
}

\newsec{Liouville interpretation of the solutions}

There is a Virasoro interpretation of each of these classes
of monodromies as well.  It is intimately related to the
mathematics of the classical solutions of Liouville theory.
Given a stress tensor $T_{uu}$, one can construct two
solutions $\psi_1$, $\psi_2$
to the Schrodinger equation with $T_{uu}$ as the potential:
\eqn\schr{
 (\d_u^2+T_{uu})\psi=0\ .
}
Setting the Wronskian of the two solutions to one, the quantity
$\FF(u)=\psi_1/\psi_2$ is a uniformizing coordinate for a 
Riemann surface, and
\eqn\lisol{
  \psi_1=\frac{\FF}{\sqrt{\d \FF}}\qquad\qquad\psi_2=\frac1{\sqrt{\d \FF}}\ .
}
The stress tensor is simply the Schwarzian derivative of $\FF$:
\eqn\schwarzian{
  T_{uu}=-\frac k2\{\FF,u\}
	=-\frac k2\Bigl[\frac{\d^3\FF}{\d\FF}
		-\frac32\Bigl(\frac{\d^2\FF}{\d\FF}\Bigr)^2\Bigr] \ .
}
Around the singularities of $T_{uu}$ 
(for example, $h(e^{iu}-e^{iu_0})^{-2}$ for a primary field), 
one has \sltwo\ monodromy
\eqn\monod{
  \pmatrix{\psi_1\cr \psi_2}(u+2\pi)=\pmatrix{a&b\cr c&d}
	\pmatrix{\psi_1\cr \psi_2}(u)\quad,\qquad ad-bc=1\ .
}
Finally, having solved for $\FF(u)$ as well as 
its counterpart $\tilde\FF(v)$ built from $T_{vv}$,
a classical solution to Liouville theory can be constructed
as $exp[-\varphi_{\sst L}]=\psi_1{\tilde\psi_2}-\psi_2{\tilde\psi}_1$
(equivalently, $\varphi=\d\log\psi_2$ is a free field which is the
Backlund transform of the Liouville field).
The three different conjugacy classes of \sltwo\
give rise to three distinct classes of solutions
to Liouville theory:
\medskip
\vbox{
\settabs\+&\sltwo\ Conj. class\qquad& Uniformizing coord.\qquad&
Riem. surf. feature\qquad& 2+1d interp.\qquad&\cr
\medskip
\+&\sltwo\ {\it Conj. class}&{\it Uniformizing coord.}&
{\it Riem. surf. feature}&2+1d {\it interp.}\cr
\+&Elliptic&$\FF(u)=\tan[\alpha u]$&
conical singularity&ptcle in $AdS_3$\cr
\+&Parabolic&$\FF(u)=a u$&cusp singularity&extreme BH\cr
\+&Hyperbolic&$\FF(u)=\exp[\lam u]$&handle&nonextreme BH\cr
}
\bigskip

\noindent
The stress tensor is $T_{uu}=-\frac{k}{4}\alpha^2$ 
in the elliptic case;
$T_{uu}=0$ in the parabolic case; and 
$T_{uu}=\frac{k}{4}\lam^2$
in the parabolic case.
To be completely explicit, the asymptotic behavior of the
Chern-Simons connection is determined in terms of the above
data via \pure, with 
$g(u)$ the Wronskian matrix of $\psi_1$, $\psi_2$.

There is a bound $m\ell<k/2$ on particle masses; an object of
larger mass is a black hole.%
\foot{Strictly speaking, all of the considerations so far have
involved only the asymptotic behavior of the metric; one might wonder
whether a metric which is asymptotic to the BTZ solution
can match onto a nonsingular interior,
say a 2+1 stellar equilibrium solution.  The work of \cz\
shows that such solutions may exist, as well as
the existence of an upper bound on their mass.  Thus, even though 
there is no long-range gravitational force, the `attraction' of 
geodesics in $AdS_3$ causes any sufficiently high mass state
to evolve to a black hole.}
A `stringy exclusion principle' has been proposed \refs{\maldastrom}
to explain the truncation of the BPS
supergravity spectrum.
This latter bound amounts to $m\ell <k$; it would seem, therefore, that
the bound is inextricable from black hole physics%
\foot{The mismatch between these two bounds might be due
to the considerations of the previous footnote.},
and that one should not expect to be able to see it
in perturbative supergravity
(or perturbative string theory, for that matter).

We see that to each classical solution of 2+1 gravity, the 
asymptotic behavior of the metric at infinity is associated uniquely with
a classical Liouville field; in turn, this Liouville field
is in one-to-one correspondence with the stress tensor data
of the dual conformal field theory.

Of course, a given asymptotic behavior can match onto
many interior solutions; for instance, 
all stationary multiparticle states
with a given total ADM energy will asymptote to \ptcleconn.
We can qualitatively build
such multiparticle states in gravity in the context of
the AdS/CFT correspondence.
Geometrically, particles are Wilson lines if they are
heavy enough (and field wavepackets if they are light), and 
gravitational waves are distortions of the geometry which can be introduced
separately at each source (by cutting out a solid cylinder at fixed 
small radius enclosing the source, performing a conformal
transformation on the boundary of the cylinder, and gluing
the geometry back together).
On the CFT side, the operators related to matter
fields in the low-energy bulk theory are scaling operators
of small anomalous dimension (relative to $k/4$)
\refs{\gkp,\wittenads}.
The primary state $\ket\alpha=\OO_\alpha\ket{0}$ 
and its \sltwo\ descendants 
$L_{-1}^m {\bar L}_{-1}^{\bar m}\ket\alpha$ form a basis
of modes of the matter field \refs{\bkl,\deboer,\bdhm,\bklt},
while the higher Virasoro raising operators 
($L_{-n},{\bar L}_{-n}$, $n\ge2$)
acting on these states amount adding a 
`gravitational wave' along the lines of \adswave.
In the semiclassical limit of large $k$, with well-separated
sources, one should be able to largely ignore the nonlinear 
effects of gravity and treat the sources as independent.
The analogous CFT state would be
\eqn\multiptcl{
  \prod_\alpha \Bigl(\prod_{\{n_i,{\bar n}_{\bar i}\}}
	L_{-n_i}{\bar L}_{{\bar n}_{\bar i}}\OO_\alpha\Bigr)\ket{0}\ .
}
Once the black hole transition is reached at $\lz=\lzb=k/4$,
the states of this form are a highly redundant description
of the Hilbert space (\cf\ \bdhm\ for a discussion); 
this is a manifestation of the `holography' (vastly reduced number
of degrees of freedom relative to local quantum field theory)
of the construction.  We will return to this subject below.

The fact that Euclidean saddle points of the gravitational action 
encode the black hole density of states \refs{\gibhawk}, 
in a manner that does not admit interpretation in terms of state counting,
supports the present perspective -- that gravity is thermodynamic
in nature and should not know about the microphysics.
Chern-Simons/Liouville theory codes the 2+1 black hole density of states
as the action of a saddle point in the Euclidean domain
\refs{\carteit,\bbo,\maldastrom,\banmen}.
It is in this way that the classical central charge \centext\
of Liouville theory enters the discussion.
The Euclidean continuation of the BTZ solution \bhconn\
is a solid torus (a disk times a circle).  
The parameters of \bhconn\ continue to 
$z_+=z_-^*=r_++i\alpha$ (\ie\ $r_-=i\alpha$), and $u=-v^*$.
The boundary of this space is a two-dimensional
Euclidean torus of modulus $\tau=i/z_+$.
This periodicity is required in order that the 
coordinate map $\FF(u)=\exp[z_+u]$ (which determines
the classical Liouville field) is single-valued
under $u\rightarrow u+2\pi\tau$.
Note that this coordinate map is related
to the thermal nature of the corresponding CFT state;
we will see shortly that it is directly related
to the Cardy formula \refs{\cardy} for the density of states in
conformal field theory.
Equivalently, the map $u\rightarrow\FF(u)$ in the CFT induces
a Bogoliubov transformation on the field modes which
generates a thermal density matrix, and thus
relates the Minkowski and Rindler vacua in 1+1 dimensions.
The inverse temperature generated
by the transformation is $\beta=2\pi\,{\rm Im}\tau$.
The Liouville zero-mode momentum is $\d_u\varphi=z_+$.
The classical Liouville action on this torus is thus
\eqn\liact{\eqalign{
  \II_{\rm cl}=&~\frac k{4\pi} \int d^2u\;|\d_u\varphi|^2
	=\frac k2\cdot 2\pi{\rm Im}\tau\cdot|z_+|^2\cr
	=&~\frac{2\pi r_+}{8G}=\beta(M+\Omega J)-S
}}
The last line is the standard Gibbons-Hawking result \refs{\gibhawk}, 
specialized to 2+1 BTZ black holes 
(\cf\ \refs{\carteit,\bbo,\maldastrom,\banmen});
here $\Omega=-\alpha/r_+$ is the angular potential,
and the mass $M$ and angular momentum $J$ are given in \bhenspin.

There is a direct relation between the preceding determination
of the gravitational entropy from Liouville theory and the Cardy 
formula for the density of states of a unitary CFT; 
both arise from the anomalous transformation law
of the stress tensor (see equation \schwarzian)
\eqn\Tanom{
  T(w)dw^2=T(z)dz^2+\frac{c}{12}\{z,w\}dw^2\ .
}
We follow closely the analysis of \refs{\dansteve}.
The partition function of a conformal field theory is a
section of a projective line bundle $E_c$
on the moduli space $\MM_g$ of Riemann surfaces.  
The projective connection $\AA$ is determined 
from the CFT stress tensor by integration
against a Beltrami differential $\mu$
\eqn\Zderiv{
  \frac{1}{2\pi i}\int d^2z \;T(\bar m,m,z)\mu(z,\zbar)=
	\ZZ^{-1}(\delta_\mu\ZZ)\ .
}
Let $\AA=0$ in coordinates $w\sim w+m\tau+n$ on the torus.
This coordinate chart does not extend to the 
$-1/\tau\rightarrow i\infty$ boundary of the moduli space $\MM_1$;
good coordinates there are $z=\exp[2\pi i w/\tau]$, 
$q=\exp[-2\pi i/\tau]$.  This coordinate transformation is 
the same as that of the classical Liouville solution.
The anomalous transformation law of the stress
tensor \Tanom\ induces the transformation of the partition
function \refs{\dansteve} via \Zderiv 
\eqn\Zanom{
  \tilde\ZZ(z,q)=(q\bar q)^{c/24}\ZZ(w,\tau)\ ,
}
and the LHS is regular as $q\rightarrow 0$.  
The asymptotic behavior of the density of states then follows
by the saddle point approximation
\eqn\saddle{\eqalign{
  \exp[S(h,\bar h)]=&~\frac{1}{(2\pi i)^2}
		\int\frac{dqd\bar q}{q^{h+1}\bar q^{\bar h+1}}\ZZ(w,\tau)\cr
	\sim&~\int d\tau d\bar\tau\;\exp\Bigl[-2\pi i(h\tau-\bar h\bar\tau)
		+2\pi i\Bigl(\frac{1}{\tau}-\frac{1}{\bar\tau}\Bigr)
		\frac{c}{24}\Bigr]\tilde\ZZ(z,q)\cr
	\sim&~ \exp\bigl[2\pi\bigl((\coeff16 ch)^\half+
		(\coeff16 c\bar h)^\half\bigr)\bigr]
		\tilde\ZZ\bigl(q=e^{-2\pi ({24h}/{c})^{1/2}}\bigr)\ ;
}}
the factor $\tilde\ZZ$ of the last line is slowly varying at 
its point of evaluation.  With the identifications 
\eqn\thermodata{\eqalign{
	h=\hf(\ell M+J)\quad,& \qquad \beta=2\pi\,{\rm Im}\tau\cr
	\bar h=\hf(\ell M-J)\quad,&\qquad
		\Omega=-\coeff{{\rm Im}\tau}{{\rm Re}\tau}\ ,
}}
one recovers the free energy as $F=M+\Omega J-S/\beta=\II_{\rm cl}/\beta$.
Thus the saddle point that controls the high energy density
of states of a unitary conformal field theory is
the same as the Liouville saddle point \liact.

\newsec{Black hole thermodynamics}

Does 2+1d gravity itself provide an accounting
of BTZ black hole microstates? No.
The dynamics of the Chern-Simons gravity theory is
completely equivalent to that of the boundary Liouville theory;
the bulk theory is pure gauge.
As discussed in \adsmm, the asymptotic black hole level density
\levdens, where $c=6k$, is {\it not}
the level density of the Liouville theory, whose
spectrum is \levdens\ with $c_{\rm eff}=1$ \kutseib.
The point is that the Chern-Simons/Liouville theory
is an {\it effective} description.  As we see from \collect,
the gravitational degrees of freedom are 
{\it collective coordinates}
of the underlying microphysics that capture the properties of the
{\it current sector} of the CFT dynamics
(\ie\ the Verma module of the identity).

The Hilbert space of the Liouville theory has in it
only the density of states of a single scalar field.
The Liouville field precisely accounts
for the asymptotic data of 2+1 gravity.  The center-of-mass
mode of the Liouville field measures the holonomy
of the Chern-Simons connection, as shown above.
The Liouville oscillators encode the `gravitational wave' data 
-- the non-constant modes of $T_{uu}$, $T_{vv}$ \listress\ --
that one can add to a given solution.
An example of this is the wave solution \adswave.
The set of oscillator states of a single free boson
is in one-to one correspondence with
the generators of the Virasoro algebra;
this fact is the basis of the string no-ghost theorem 
(\cf\ \refs{\thorn} and references therein).
Thus $c_{\rm eff}=1$ describes the density of states of
pure gravity.

Because the current sector is universal and couples to
all states of the CFT/gravity theory, it takes into account
all of the degrees of freedom of the microphysics;
however, it cannot distinguish microstates 
with the same asymptotic metric (\eg\ the same energy).
The current sector is thus thermodynamic in character.
The currents are the generating functions of Noether charges 
on the gravity side --
precisely the objects that one couples to thermodynamics
via the introduction of a set of conjugate potentials.
One sees this explicitly in the way the saddle point
action \liact\ measures the density of states.
Extremizing the classical (Euclidean) gravitational action
amounts to extremizing a thermodynamic function,
and determines properties of the equilibrium state.

It is thus incorrect to try to count microstates by quantizing
the bulk gravity/Liouville theory; gravity fluctuations
are not given by Liouville fluctuations (these have nothing
a priori to do with the fluctuations of the full CFT%
\foot{A useful analogy might be fluid dynamics; it is
in general incorrect to quantize the Euler equations
(which are in any event highly nonlinear and nonrenormalizable),
rather one quantizes the underlying many-body problem
and then introduces collective variables suitable 
for the long-wavelength limit of the quantum system.
In other words, quantum fluid-dynamics is not necessarily
quantum-fluid dynamics.}),
although of course all pure stress tensor correlators
will agree as a consequence of the Virasoro
Ward identities.  The Liouville theory also
characterizes the conformal properties 
of thermal states in the conformal field theory,
as shown above.

Another class of calculations  
\refs{\carlipent,\dunno}
attempts to localize gravitational entropy in a set of
degrees of freedom on the black hole horizon.
One problem with this field-theoretic
approach is that gravitational entropy is universal
(this is the main strength of the Brown-Henneaux/Strominger construction),
whereas any attempt to localize microstates
on the horizon depends strongly on nonuniversal
properties of the theory -- how much supersymmetry it has,
what the matter content is, \etc\ \  
In Lorentz signature,
quantitative calculations of this sort have only been 
performed \carlipent\ in the pure gravity theory,
and rely strongly on the fact that
the dimension of the gauge group $SL(2,R)\times SL(2,R)$
in \threedgrav\ is six.  
If one modifies the theory,
for instance enlarging the gauge group to that
of extended Chern-Simons supergravity, the prescription of 
\carlipent\ no longer reproduces the black hole entropy \adsmm.
A number of other difficulties inherent to this approach have recently
been elaborated \dunno.
Essentially, there is no canonically defined CFT on the horizon
due to the many possible choices of boundary condition
and ways of embedding the Virasoro algebra into the 
three-dimensional diffeomorphism group.
Even if these difficulties could somehow be resolved,
the addition of dynamical matter gives a further (divergent) 
contribution \tangle\ to the entropy. 
This additional contribution is
entanglement entropy of field modes on either
side of the horizon, and is not identical
to information content and thermodynamic entropy.
Naively one would expect the entanglement entropy 
to depend on the number and kind of matter
fields contributing to it.
Each time one considers a different
matter content, one faces the task of explaining why the
black hole entropy remains unchanged. 

The standard field-theoretic approach to quantum gravity
and black holes takes locality as a given.  One is then led
to ascribe a physical reality to the black hole horizon
as a boundary of causal contact, and
a certain degree of separability or distinctness
between degrees of freedom inside and outside this horizon.
This assumption leads directly to the black hole
information paradox (\cf\ \fpst).
However, in the classical theory the position of the horizon
is not locally defined (the local notion of apparent horizon
is not directly related to signal propagation); and in the
quantum theory, the position of the horizon is not well-defined
due to quantum fluctuations in the geometry.
The very presence of a boundary at the horizon
violates the reparametrization
gauge constraints, since the horizon degrees of freedom
that are counted in \refs{\carlipent,\dunno}
are `would-be' gauge transformations that are not symmetries.
(On the other hand, the analogous gauge transformations 
on the boundary at infinity are self-consistently frozen
because fluctuations are suppressed by the infinite volume 
of the asymptotic region; imposition of a fixed classical geometry
is thus sensible.)  The above discussion
suggests one should interpret
the horizon gauge transformations as analogues of
the descendants of particle states \multiptcl,
in which case they grow asymptotically as $c_{\rm eff}=\OO(1)$ 
rather than as $c=6k$.

The recent constructions of black holes
in M/string theory (in the context of both the Maldacena conjecture
and Matrix theory) have the property that any attempt
to concentrate too much energy in a given region
(as measured by stationary observers from afar) 
results in nonlocality on that scale, \eg\ in an 
interdependence of creation operators of modes in the
low-energy effective field theory 
(\cf\ \bdhm\ for a recent discussion in the context of the
AdS/CFT correspondence).
One sees this in matrix theory black holes \matbh, where
the size of the horizon is the uncertainty bound on
the quantum wavefunction of the matrix degrees of freedom;
and in AdS-Schwarzschild black holes of the AdS/CFT correspondence, 
where the horizon scale corresponds
to the thermal wavelength in the conformal field theory
\refs{\maldaconj,\wittenads,\bdhm,\bklt}.
In either case, any observable is unavoidably entangled strongly
with the black hole degrees of freedom at this scale;
this is a reflection of the fact that,
at the length scale $r_+$ characteristic of the black hole,
the number of independent available degrees
of freedom is much smaller than in local field theory.
In particular, there is no separation of degrees of freedom
`inside' and `outside' of a black hole, and therefore 
no sense to the erection of a boundary that separates
the black hole interior from the rest of spacetime.
The correct classical limit then has only the boundary
at spatial infinity of anti-de Sitter space; this justifies
the above treatment of the 2+1 gravity dynamics,
where only the Liouville degrees of freedom at infinity
were considered.

Most of our discussion has ignored the 
effects of matter fields coupled to 2+1 gravity. 
Even though gravity itself appears to lack the necessary 
degrees of freedom to account for black hole entropy,
might it be possible to enumerate a set
of gravitationally dressed matter fields that
reproduce the correct answer?  
In the context of perturbative local field theory,
this approach seems to lead to the divergent entanglement
entropies mentioned above.  A more promising approach
\refs{\stromtalk}
might be to enumerate a basis for the Hilbert space of
the dual CFT of the AdS/CFT correspondence,
using the creation operators of matter particles 
\refs{\maldastrom,\deboer,\bdhm,\bklt} as in \multiptcl.  
However, precisely in the regime of energy levels of
interest $\lz\sim k$ and above, the nature of the Hilbert space changes
character, and the products of such operators become a highly 
overcomplete basis \bdhm;
this is simply a restatement of `holography' \susswit, or
of the `stringy exclusion principle' of \maldastrom.  
Rather than having mode creation operators spanning a free algebra,
as one supposes in local quantum field theory, the algebra
satisfies additional relations.  These relations
are manifested in the operator product expansion of the dual CFT, 
which forces the creation modes of different particle states
to be interdependent at next-to-leading order in the $1/k$ expansion.
The interdependence of the creation modes has dramatic effects;
for example \maldastrom, the \nth\ power ($n>2k$) of a supergraviton
creation operator is not an $n$-particle state (this concept 
is well-defined for BPS states),
rather it is a current algebra descendant of a state
with $m<2k$ particles.
Because the energy of such a state is above the
threshold to create a black hole, 
the dependencies among the creation modes 
appear to be intimately connected with the Bekenstein bound.
We might call the additional constraints
`black hole operator product relations'.
As a consequence, it is not clear that one
can assign to the generators of 
any proposed basis of `multiparticle states'
the same meaning which they have in the construction of dilute
gases of multiparticle states about the AdS vacuum
(discussed at the end of the last section).


\vskip 1cm
\noindent{\bf Acknowledgments:}
The author thanks the Institute for Theoretical Physics, Santa Barbara,
for hospitality and support during the course of this work;
and the organizers of the Amsterdam Summer Workshop,
`String Theory and Black Holes', for the opportunity
to present and discuss the results in a pleasant and stimulating environment.
This work was also supported by DOE grant DE-FG02-90ER-40560.



\listrefs
\end